\documentclass[journal,10pt]{IEEEtran}
\usepackage{multirow}
\usepackage{soul}
\usepackage[noadjust]{cite}
\usepackage{cuted}
\usepackage{amsmath,amssymb,graphicx,amsthm,mathrsfs}
\usepackage{epstopdf}
\usepackage{enumitem}
\usepackage{enumitem}

\usepackage{algorithm}
\usepackage{algorithmicx}
\usepackage{algpseudocode}
\floatstyle{plain}
\newfloat{twocolequfloat}{b}{zzz}
\floatname{twocolequfloat}{Equation}
\newcommand{\argmin}{\operatornamewithlimits{argmin}}

\newcommand{\bal}{\begin{aligned}}
\newcommand{\eal}{\end{aligned}}
\newcommand{\beq}{\begin{equation}}
\newcommand{\eeq}{\end{equation}}
\makeatletter
\newcommand{\thickhline}{%
    \noalign {\ifnum 0=`}\fi \hrule height 1.15pt
    \futurelet \reserved@a \@xhline
}
\usepackage{booktabs,caption}

\def\X{{\bf X}}
\def\Xk{{\bf X}_k}

\def\x{{\bf x}}
\def\u{{\bf u}}
\def\uk{{\bf u}_k}

\def\y{{\bf y}}
\def\Yk{{\bf Y}_k}

\def\f{{\bf{f}}}
\def\v{{\bf{v}}}
\def\Xkm{{\bf X}_{k-1}}

\def\h{{\bf{h}}}
\def\nk{{\bf n}_k}
\def\Mk{{M_k}}

\def\tr{\tilde{\bf r}}

\def\Xz{{\bf X}_0}
\def\b{{\bf b}}
\def\J{{\bf J}}

\def\bz{{\bf{0}}}

\def\su{\mathbb{U}}

\def\sR{\mathbb{R}}
\def\sB{\mathbb{B}}

\def\sig0{\sigma_0}
\def\sig1{\sigma_1}
\def\Qb{\overline{Q}}

\def\Q{{\bf{Q}}}

\def\dQ{\Delta{\bf{Q}}}

\def\k{{\bf{k}}}
\def\kt{\tilde{\bf{k}}}
\def\K{{\bf{K}}}
\def\Kt{\tilde{\bf{K}}}
\def\H{{\bf{H}}}
\def\Ht{\tilde{\bf{H}}}

\def\bu{({\bf{b}},{\bf{u}})}
\def\bu0{({\bf{b}}_0,{\bf{u}}_0)}
\def\bu1{({\bf{b}}_1,{\bf{u}}_1)}

\def\but0{(\tilde{{\bf{b}}}_0,\tilde{{\bf{u}}}_0)}

\def\but{(\tilde{{\bf{b}}},\tilde{{\bf{u}}})}

\def\c{{\bf{c}}}

\def\h{{\bf{h}}}
\def\B{{\bf{B}}}

\def\lar{\leftarrow}
\def\Phib{{\pmb{\Phi}}}

\def\N{\mathcal{N}}
\def\GP{\mathcal{GP}}
\def\D{{\bf D}}
\def\t{{\bf t}}
\def\T{{\bf T}}
\def\vt{{\bf v}}
\def\alp{{\mbox{\boldmath $\alpha$}}}
\def\Rt{\tilde{\bf{R}}}
\usepackage{relsize}

\begin{document}


\title{Control of Gene Regulatory Networks with Noisy
  Measurements and Uncertain Inputs}
\author{Mahdi Imani,~\IEEEmembership{Student~Member,~IEEE}~and~Ulisses~M.~Braga-Neto,~\IEEEmembership{Senior~Member,~IEEE}%
\thanks{M. Imani and U.M. Braga-Neto are with the Department of Electrical and Computer Engineering,
Texas A\&M University, College Station, TX 77843 USA (e-mail: m.imani88@tamu.edu, ulisses@ece.tamu.edu)}
}

\oddsidemargin 0.0in
\textwidth 6.5in

\maketitle

\vspace{-0.3cm}
\begin{abstract}
  This paper is concerned with the
  problem of stochastic control of gene regulatory networks (GRNs) observed indirectly through noisy measurements and with uncertainty in the intervention inputs. The
  partial observability of the gene states and uncertainty in the
  intervention process are accounted for by modeling GRNs using the
  partially-observed Boolean dynamical system (POBDS) signal
  model with noisy gene expression measurements. Obtaining the
  optimal infinite-horizon control strategy for this problem is not
  attainable in general, and we apply reinforcement
  learning and Gaussian process techniques to find a near-optimal
  solution. The POBDS is first transformed to a directly-observed
  Markov Decision Process in a continuous belief space, and the
  Gaussian process is used for modeling the cost function over the
  belief and intervention spaces. Reinforcement learning
  then is used to learn the cost function from the available gene
 expression data. In addition, we employ sparsification, which enables
 the control of large partially-observed GRNs. The performance of the resulting algorithm is studied through a comprehensive set of numerical experiments using synthetic gene expression data generated from a melanoma gene regulatory network. 
\end{abstract}


\section{Introduction}
\label{sec:Intro}
A key purpose of control of gene regulatory networks (GRNs) is to derive appropriate strategies to avoid undesirable states,
such as those associated with disease.
GRNs play a crucial role in every process of cellular life, including cell differentiation, metabolism, the cell cycle and signal transduction~\cite{karlebach2008modelling}. 
Several models were introduced in literature to mathematically capture
the behavior of gene regulatory networks, such as probabilistic
Boolean network (PBN)~\cite{shmulevich2002boolean}, Bayesian
networks~\cite{friedman2000using}, and Boolean control
networks~\cite{cheng2009controllability}. Several intervention
strategies were also developed for control of GRNs
(e.g. \cite{datta2003external,pal2006optimal,cheng2010linear}). 

Most of the existing approaches assume that the Boolean state of genes is
directly observable. In the current paper, the goal is to obtain
appropriate intervention strategies to beneficially alter network
dynamics, while assuming that the GRN is only indirectly observable
through noisy gene expression measurements. In addition, we assume that the
intervention input itself has uncertain effects. 
The signal model used in our approach is the partially-observable
Boolean dynamical system (POBDS) model
\cite{Brag:11,ImanBrag:17}. Several tools for POBDSs have been
developed in recent years, such as the optimal filter and smoother
based on the minimum mean square error (MMSE) criterion, called the
Boolean Kalman Filter (BKF)~\cite{Brag:11} and Boolean Kalman Smoother
(BKS)~\cite{ImanBrag:15b}, respectively. In addition, particle
filtering implementations of these filters, as well as schemes for
handling correlated noise, simultaneous state and parameter
estimation, network inference, and fault detection for POBDSs were
developed \cite{ImanBrag:PF,LeviBrag:17,ImanBrag:17,ImanBrag:15a,BahaBrag:15}. The software tool ``BoolFilter"~\cite{BoolFilter} is available under R library for estimation and identification of partially-observed Boolean dynamical systems.

In~\cite{ImanBrag:16a,ImanBrag:17b}, a state feedback controller for POBDSs is
proposed based on optimal infinite horizon control of the Boolean
state process, with the Boolean Kalman filter as state observer. This
method, which is called V\_BKF in this paper, has similarities to the
Q\_MDP method introduced in~\cite{Littman95} for a general nonlinear
state space model, which also does not employ the belief space when
obtaining the control policy.  Although this type of controller can be
effective in some domains, the obtained policies do not take informed
control action and might perform poorly in domains where repeated
information gathering is
necessary~\cite{spaan2005perseus,pineau2006anytime,ross2007aems,pineau2003point}.
In addition, the point- based value iteration method is used in~\cite{ImanBrag:16c}  to control POBDSs with finite observation spaces. However, point-based techniques are only suitable for
relatively small state
spaces~\cite{spaan2005perseus,smith2004heuristic,smith2012point,porta2006point}.
 
In this paper, we transform the partially-observed Boolean state space into
belief space, which is a continuous observed state space, and learn the
optimal policy in this space. We use the Gaussian process
as a nonparametric technique to model the cost function over both
belief and intervention spaces, and reinforcement learning is employed
to learn the cost function by collecting a finite set of samples. It
should be noted that unlike parametric representation techniques in
which the uncertainty of the cost function is encoded in the estimate
of the parameters, nonparametric Gaussian processes are Bayesian
representation of the cost function, which yields several benefits
such as:
\begin{enumerate}
\item Prior knowledge about the cost in the belief and intervention spaces can be easily used to increase the learning rate.
\item The exploration/exploitation trade-off, which is a crucial fact
  in the performance of any reinforcement learning technique, can be
  easily addressed using the notion of uncertainty that is provided by Gaussian process model. 
\item The concept of risk can be taken into account in obtaining a robust intervention strategy. 
\end{enumerate}
The above benefits will be discussed in detail throughout the text.

The article is organized as follows. In Section~\ref{sec:BDS}, the
POBDS model used in this paper is introduced. Then, the
infinite-horizon control problem is formulated in
Section~\ref{sec:Control}. In Section~\ref{sec:RL}, reinforcement
learning and Gaussian processes are used for control of
partially-observed GRNs. The sparsification technique for control of
large GRNs is discussed in Section~\ref{sec:spars}.  Results of a
numerical experiment using a melanoma gene regulatory network observed
through synthetic gene expression time series are reported and discussed in
Section~\ref{sec:NE}. Finally, Section~\ref{sec:con} contains
concluding remarks.

\section{POBDS Model}
\label{sec:BDS}

In this section, the POBDS model is briefly introduced. It consists of
a state model that describes the evolution of the Boolean dynamical
system, which includes the system input, and an observation model that
relates the state to the system output (measurements). More details
can be found in \cite{Brag:11,ImanBrag:17}.

\subsection{POBDS State Model}

Assume that the system is described by a {\em 
state process} $\{\Xk; k=0,1,\ldots\}$, 
 where $\Xk \in \{0,1\}^d$ represents the activation/inactivation state of the
genes at time~$k$. The state of the genes is affected by a sequence of
\textit{control inputs} $\{\uk; k = 0, 1, \ldots\}$, where
$\u_k\in\su=\{0,1\}^r$, $r\leq d$, represents a purposeful control
input. The states are assumed to be updated at each discrete
time through the following nonlinear signal model:
\beq
\bal
  \Xk &\,=\, \f\left(\Xkm,\u_{k-1}\oplus\vt_{k-1}\right)
  \,\oplus\, \nk \,,
\label{eq-sgnmodel}
\eal
\eeq
for $k=1,2,\ldots$, where $\f: \{0,1\}^{d}\times \{0,1\}^r\rightarrow \{0,1\}^d$ is a Boolean function called
the {\em network function}, ``$\oplus$'' indicates componentwise
modulo-2 addition, $\nk \in \{0,1\}^d$ is Boolean transition
noise, and $\vt_k$ is Boolean noise that makes the control input uncertain.
The noise processes $\{\nk; k=1,2,\ldots\}$  and $\{\vt_{k}; k=0,1,\ldots\}$ are assumed to
be ``white'' in the sense that the noise at distinct time points are
independent random variables. We also assume that noise processes are
independent of each other and independent of the initial state
$\Xz$. The way that the input influences state evolution is part of
the function $\f$; typically, as will be the case here, each bit in the input $\u_{k-1}$, if it is one, flips the value of a specified bit of the Boolean state $\Xk$. Note that, in some cases, an input bit
will not have any effect, since it may be reset by the corresponding
bit in the noise $\vt_{k-1}$.

We assume a noise distribution where the bits in 
$\nk$ and $\vt_k$ are i.i.d. (the general non-i.i.d. case can be similarly handled,
at the expense of introducing more parameters), with
$P(\nk(i) = 1) = p$ and $P(\vt_k(j) = 1) = q$, for $i = 1,\ldots,d, \,j=1,\ldots,r$.
Parameters $0 < p,q < 1/2$ correspond to the amount of ``perturbation"€
to the Boolean state and intervention processes,
respectively --- the cases $p = 1/2$ and $q = 1/2$ correspond to
maximum uncertainty. 

Let $(\x^1,\ldots,\x^{2^d})$ and $(\v^1,\ldots,\v^{2^r})$ be arbitrary enumeration of the
possible state and intervention noise vectors.
The \textit{prediction matrix} is the transition matrix of
the underlying controlled Markov chain, given by:
\beq\label{eq-M}
\bal
   &(\Mk(\u))_{ij} \,=\, P(\Xk = \x^i \mid \Xkm = \x^j,\u_{k-1}=\u)\\
   &=\,\sum_{s=1}^{2^r} P(\Xk \!= \x^i \mid \Xkm \!=
   \x^j,\u_{k-1}\!=\u,\v_{k-1}\!=\v^s)\\[-1ex]
   &\qquad\qquad\qquad \quad\times P(\v_{k-1}=\v^s)\\
   & =\,\sum_{s=1}^{2^r} q^{||\u\,\oplus\,\v^s||_1}(1\!-\!q)^{r-||\u\,\oplus\,\v^s||_1}p^{||\f(\x^j,\u\,\oplus\,\v^s) \,\oplus \,\x^i||_1} \\[-1ex]
   &\qquad\qquad\qquad \quad\times (1\!-\!p)^{d-||\f(\x^j,\u\,\oplus\,\v^s) \,\oplus\, \x^i||_1}\,,
   \eal
\eeq
for $i,j = 1,\ldots,2^d$ and given $\u \in \su$. 

\subsection{POBDS Observation Model}


In this paper, we assume a POBDS observation model that corresponds to
Gaussian gene expression measurements at each time point.  This is an
appropriate model for many important gene-expression measurement
technologies, such as cDNA microarrays \cite{ChenDougBitt:97} and live
cell imaging-based assays \cite{Huaetal:12}, in which gene expression
measurements are continuous and unimodal (within a single population of
interest).

Let $\Yk = (\Yk(1),\ldots,\Yk(d))$ be a vector containing the measurements at time $k$, for $k = 1,2,\ldots$. The component
$\Yk(j) \in R$ is the abundance measurement corresponding to
transcript $j$, for $j = 1,\ldots,d$. 
We assume conditional independency of the
measurements given the state as:
\beq
\bal
P(\Yk &= \y \mid \Xk=\x)\,\\
&=\,\prod_{j=1}^d P(\Yk(j) = \y(j) \mid \Xk(j)=\x(j)),
\eal
\label{eq:ind}
\eeq
and adopt a Gaussian model,
\beq
\label{eq:NGS1}
\bal
P(\Yk(j) =& \y(j) \mid \Xk(j)=\x(j))\\
 &=\frac{1}{\sqrt{2\,\pi\,\sigma^2_j}}\, \exp\left(-\frac{(\y(j)-\mu_j)^2}{2\,\sigma_j^2}\right),
\eal
\eeq
where $\mu_j$ and $\sigma_j>0$ are the mean and standard deviation of
the abundance of transcript $j$, respectively, for $j=1,\ldots,d$. 

According to the Boolean state model,
there are two possible states for the abundance of transcript $j$:
high, if $\x(j) = 1$, and low, if $\x(j) = 0$. 
Accordingly, we model $\mu_{j}$ and $\sigma_j$~as:
\beq
\bal
\mu_{j}\,=\,\mu^0_{j}\,\left(1-\x(j)\right)\,+\,\mu^1_{j}\,\x(j)\,,\\
\sigma_{j}\,=\,\sigma^0_{j}\,\left(1-\x(j)\right)\,+\,\sigma^1_{j}\,\x(j)\,,
\label{eq:NGS2}
\eal
\eeq
where the parameters $(\mu_j^0,\sigma_j^0>0)$ and
$(\mu_j^1,\sigma_j^1>0)$ specify the means and standard deviations
of the abundance of transcript $j$ in the inactivated and activated states, respectively.

Based on equations~(\ref{eq:ind}), (\ref{eq:NGS1}) and (\ref{eq:NGS2}), the \textit{update
  matrix}, which is a diagonal matrix of size $2^d\times 2^d$, is given by:
\beq
\bal
&\left(T_k(\y)\right)_{ii} \,=\, P\left(\Yk = \y \mid \Xk = \x^i\right)\\
&=\left(\prod_{j=1}^d \frac{1}{\sqrt{2\,\pi\,\left(\sigma^0_{j}\,(1-\x^i(j))\,+\,\sigma^1_{j}\,\x^i(j)\right)^2}}\right)\, \\
&\times\exp\left(-\mathlarger{‎‎\sum}_{j=1}^d\frac{\left(\y(j)-\mu^0_{j}(1-\x^i(j))-\mu^1_{j}\x^i(j)\right)^2}{2\,\left(\sigma^0_{j}(1-\x^i(j))+\sigma^1_{j}\x^i(j)\right)^2}\right),\\
\eal\label{eq:GRN5}
\eeq
for $i=1,\ldots,2^d$ and observed $\y \in R^d$. Typical values for the parameters are
given in Section~\ref{sec:NE} when we discuss the numerical experiments
performed to evaluate the proposed approach.

\section{Infinite-Horizon Control}
\label{sec:Control}

In this section, the infinite-horizon control problem for the POBDS
model is formulated. The goal of control in this paper is to select
the appropriate external input $\u_k\in\su$ at each time $k$ to make
the network spend the least amount of time, on average, in undesirable
states (e.g., states corresponding to cell proliferation,
which are undesirable, as they may be associated with
cancer~\cite{Wein:06}).

For the infinite-horizon control problem, we assume that the system
prediction matrix $M_k(\u)$ and update matrix $T_k(\y)$ can only
depend on time through the control input $\u \in \su$ and measurement
$\y \in R^d$, respectively. We will thus drop the index $k$ and write
simply $M(\u)$ and $T(\y)$.

Since the state of the system is not observed directly, all available for decision making at each time step are the observations up
to current time $\y_{1:k}=(\y_1,\ldots,\y_k)$, and the control input applied to the
system up to previous time step
$\u_{0:k-1}=(\u_1,\ldots,\u_{k-1})$. Rather than
storing the history of observations and control inputs, we record
the probability of states given that information at each time
step. This probability distribution is known as the \textit{belief
  state} at time $k$, given~by:
\beq\label{eq:bk}
\b_k(i)\,=\, P(\Xk=\x^i \mid \y_{1:k},\u_{0:k-1})\,,
\eeq
for $i=1,\ldots,2^d$. The initial belief state
is simply the initial state distribution, $\b_0(i)=P(\X_0=\x^i)$, for
$i=1,\ldots,2^d$. Since $0\leq \b(i)\leq 1$ and
$\sum_{i=1}^{2^d} \b(i)=1$, a belief vector $\b_k$ is a point in a $(2^{d}-1)$-dimensional
simplex $\sB$, called the \textit{belief space}.

Assuming $\b$ is the current belief state of the system, if the
control input $\u$ is applied and observation $\y$ is made, the new
belief can be obtained by using Bayes' rule as:
\beq\label{eq:belief2}
\b^{\u,\y}\,=\,\frac{T(\y)\,M(\u)\,\b}{\Vert T(\y)\,M(\u)\,\b\Vert_1}\,,
\eeq
where $\Vert \cdot\Vert_1$ denotes the $L_1$-norm of a vector. Thus, by
using the concept of belief state, a POBDS can be transformed into a Markov decision process (MDP)
with a state transition probability in the belief space $\sB$, given~by:
\beq\label{eq:pb}
p(\b'\mid \b,\u)\,=\,\int_{\y\in \sR^d}\,\Vert T(\y)\,M(\u)\,\b\Vert_1\,I_{\b'=\b^{\u,\y}}\,d\y,
\eeq
where $I_{\b'=\b^{\u,\y}}$ is an indicator function which returns 1 if $\b'=\b^{\u,\y}$ and 0 otherwise. 

Now, let $c(\x^i,\u)$ be a bounded cost
of control for state $\x^i$ and control input $\u$, for
$i=1,\ldots,2^d$ and $\u\in\su$. 
The cost can be transformed to belief space as follows:
\beq
  g(\b,\u)\,=\,\sum_{i=1}^{2^d} c(\x^i,\u)\,\b(i)\,.
\eeq
The goal of infinite-horizon control is to minimize the following cost function by choosing the appropriate control input at each time step:
\beq
J_{\infty}\,=\,E\left[\sum_{k=1}^{\infty} \gamma^{k}\,g(\b_k,\u_k) \,\bigg|\, \b_0\right]\,,
\eeq
where $\b_0$ is the known initial belief state, and the discount
factor $\gamma$ places a premium on minimizing the costs of early
interventions as opposed to later ones, which is sensible
from a medical perspective~\cite{pal2006optimal}.
The classical results proved in~\cite{bertsekas1995dynamic} for MDPs
can be used here. For an infinite-horizon control problem with discount
factor $\gamma$, the Bellman operator for the belief space $\sB$ can be
written as follows:
\beq\label{eq:jF1}
\bal
& T[\J](\b)\,=\,\min_{\u \in \su}  \bigg[g(\b,\u)+\gamma\,\int_{\b'\in \sB}\,p(\b'\mid \b,\u)\, \J(\b')\bigg]\,\\
&=\min_{\u \in \su}  \bigg[g(\b,\u)+\gamma\int_{\y\in \sR^d} \Vert T(\y)M(\u)\b\Vert_1\, \J(\b^{\u,\y})d\y\bigg].
\eal
\eeq
However, since the belief $\b$ is in the $(2^d-1)$-dimensional simplex
$\sB$, computing the bellman operator in~(\ref{eq:jF1}) for all belief
points is not possible. 

\section{Control Using Reinforcement Learning and Gaussian Process}
\label{sec:RL}

\subsection{Q-function as a Gaussian Process}
In this section, the cost function over belief and intervention spaces is modeled
using a Gaussian process. 
A {\em policy} is a function $\pi: \sB \rightarrow \su$, which associates a
control input to each belief state. Given a policy $\pi$, the discounted return for time step $k$ can be defined as: 
\beq\label{eq:Cpi}
C^\pi_k\,=\,\sum_{m=0}^\infty\,\gamma^m\, g(\b_{k+m+1},\u_{k+m+1})\,,
\eeq
where $C_k^\pi$ is the total accumulated cost obtained over time
following policy $\pi$. Note that $C^\pi_k$ can be written in a recursive fashion as:
\beq\label{eq:rec}
C^\pi_k\,=\,g(\b_{k+1},\u_{k+1}) + \gamma\,C^\pi_{k+1}\,.
\eeq
Due to the stochasticity in belief transition, which arises from stochasticity of state, observation and intervention processes, the discounted return is a random variable which can be decomposed into a mean $Q^\pi(\b,\u)$ and a residual $\Delta Q^\pi(\b,\u)$, for $\u\in\su$, as:
\beq\label{eq:del}
C^\pi_k(\b_k=\b,\u_k=\u)\,=\,Q^\pi(\b,\u)+\Delta Q^\pi(\b,\u)\,,
\eeq
where 
\beq
Q^\pi(\b,\u)\,=\,E_\pi[C^\pi_k\mid\b_k=\b,\u_k=\u]\,,
\eeq
where the expectation is taken over
all possible successor belief state sequences that can be observed. Notice that
the mean and residual of the return are assumed to
be independent of $k$. 

Replacing (\ref{eq:del}) into (\ref{eq:rec}), the immediate cost can be written as:
\beq
\bal
& g(\b_{k+1},\u_{k+1})\,=\,Q^\pi(\b_{k},\u_{k})-\gamma\,Q(\b_{k+1},\u_{k+1})\\
& \quad\quad +\Delta Q^\pi(\b_{k},\u_{k})-\gamma\,\Delta Q^\pi(\b_{k+1},\u_{k+1})\,.
\eal
\eeq

Let $\B_k = [(\b^0,\u^0),\ldots,(\b^k,\u^k)]^T$ be the sequence of
observed belief states and taken interventions between time steps $0$
and $k$, under policy $\pi$, we have:
\beq
 \bal
 &g(\b_1,\u_1)\,=\,Q^\pi(\b_0,\u_0)-\gamma\,Q^\pi(\b_1,\u_1)\\
 & \quad\quad +\Delta Q^\pi(\b_0,\u_0)-\gamma\,\Delta Q^\pi(\b_1,\u_1),\\
  & g(\b_2,\u_2)\,=\,Q^\pi(\b_1,\u_1)-\gamma\,Q^\pi(\b_2,\u_2)\\
 & \quad\quad +\Delta Q^\pi(\b_1,\u_1)-\gamma\,\Delta Q^\pi(\b_2,\u_2),\\
&\vdots \\
 &g(\b_k,\u_k)\,=\,Q^\pi(\b_{k-1},\u_{k-1})-\gamma\,Q^\pi(\b_k,\u_k)\\
 & \quad\quad +\Delta Q^\pi(\b_{k-1},\u_{k-1})-\gamma\,\Delta Q^\pi(\b_k,\u_k).\\
 \eal
 \eeq
The above equation can be written in a more compact form as~\cite{engel2005reinforcement}: 
\beq\label{eq:GPR}
\c_k\,=\,{\bf{H}}_k\,\Q_k^\pi+{\bf{H}}_k\,\Delta\Q_k^\pi,
\eeq
where
\beq
\bal
{\bf{c}}_k\,&=\,[g(\b_1,\u_1),\ldots,g(\b_k,\u_k)]^T\,,\\
\Q^\pi_k\,&=\,[Q^\pi(\b_0,\u_0),\ldots,Q^\pi(\b_k,\u_k)]^T\,,\\
\Delta\Q^\pi_k\,&=\,[\Delta Q^\pi(\b_0,\u_0),\ldots,\Delta Q^\pi(\b_k,\u_k)]^T\,,\\
{\bf{H}}_k\,&=\,
\begin{bmatrix}
1 & -\gamma & \ldots & 0 & 0\\
0 & 1 & \ldots & 0 & 0\\
\vdots  &  &  &   & \vdots \\
0 & 0 & \ldots & 1 & -\gamma\\
\end{bmatrix}\,.
\eal
\eeq

Due to the changes in the policy $\pi$ during the learning process, which will be addressed later in this section, $Q^\pi(\b,\u)$ is a random variable.
In order to specify a complete probabilistic generative model
connecting Q-function and costs, one needs to define a prior
distribution for the Q-function and the distribution of $\Delta
\Q$. 
A Gaussian process is a stochastic process 
which allows the extension of multivariate Gaussians to infinite-sized collections of real valued variables~\cite{rasmussen2006gaussian}.
In this paper, we use Gaussian processes for non-parametric Bayesian representation of our cost function. 
The prior distribution of the Q-function is defined as: 
\beq
Q^\pi(\b,\u)\,=\,\GP\left(\bz,k\left((\b,\u),(\b,\u)\right)\right)\,,
\eeq
where $k(.,.)$ is a real-valued kernel function over both belief and
intervention spaces.  In addition, we assume the residual $\dQ$ is
generated independently from a zero mean Gaussian distribution as
$\Delta Q^\pi(\b,\u)\sim \mathcal{N}(0,\sigma_q^2)$, where the variance
$\sigma_q^2$ is to be determined.

The kernel function $k(.,.)$ encodes our prior
beliefs on correlations between different points in belief and intervention spaces.
We consider kernels that decompose over the belief state and intervention space as:
\beq\label{eq:fac}
k\left((\b,\u),(\b',\u')\right)\,=\,k_B\left(\b,\b'\right)\,k_U\left(\u,\u'\right)\,.
\eeq
We employ the direct probabilistic representation of our intervention process in defining the kernel function in the intervention space as:
\beq 
k_U(\u,\u')=
q^{||\u\,\oplus\,\u'||_1}\,(1-q)^{r-||\u\,\oplus\,\u'||_1} \,.
\eeq 
Notice that taking control input $\u$ consecutively affects the cost function associated to all different $\u'\in \su$, for any $0<q\leq0.5$ where $q$ is the intensity of Bernoulli intervention process.

For the belief state kernel, we consider the well-known exponential kernel function:
\beq\label{eq:kernel}
k_B(\b,\b')\,=\,\sigma_f^2\,\exp\left(-\frac{||\b-\b'||^2}{2\,l^2}\right),
\eeq
where $\sigma_f^2$ determines the prior variance and $l$ denotes the
correlation at different belief points (the large values of $l$ model
more correlation of Q-function in the belief space). The parameters
$\sigma_f^2$ and $l$ are to be determined.

It is worth mentioning that factorization in equation~(\ref{eq:fac})
depends on the fact that the multiplication of two separate kernels
results in another kernel~\cite{ghoreishi2016uncertainty,scholkopf2002learning,ghoreishi2016compositional}.

Using the above assumptions, the posterior distribution of $Q^\pi(\b,\u)$ in equation~(\ref{eq:GPR}) can be obtained as~\cite{rasmussen2006gaussian,friedman2017quantifying}:
\beq\label{eq-Q}
Q^\pi(\b,\u)\mid {\bf{c}}_k,{\bf{B}}_k \sim
\N\left(\Qb(\b,\u),\text{cov}\left((\b,\u),(\b,\u)\right)\right),
\eeq
where
\beq
\bal
 & \Qb(\b,\u)\,=\,\k_k(\b,\u)^T \H_k^T(\H_k \K_k \H^T_k+\sigma_q^2 \H_k \H_k^T)^{-1}\,\c_k,\\[1ex]
&\text{cov}((\b,\u),(\b,\u))\,=\,k((\b,\u),(\b,\u))-\k_k(\b,\u)^T
\H_k^T\\
&\hspace{2ex}(\H_k\K_k\,\H_k^T+\sigma_q^2 \H_k \H_k^T)^{-1} \H_k \k_k(\b,\u)\,,
\eal
\eeq
with
\beq
\bal
\k_k(\b,\u)&=[k((\b_0,\u_0),(\b,\u)),\ldots,k((\b_k,\u_k),(\b,\u))]^T\!,\\
\K(\b,\u)&=[\k_k(\b_0,\u_0),\ldots,\k_k(\b_k,\u_k)]^T\!.
\eal
\eeq

Using the above formulation, the Q-function before observing any data
is a zero-mean Gaussian process with covariance $k((\b,\u),(\b,\u))$,
while at time step $k$, this posterior can be obtained based on the
sequence of costs $\c_k$ and sequence of observed beliefs and
interventions $\B_k=[(\b_0,\u_0),\ldots,(\b_k,\u_k)]$ using
equation~(\ref{eq-Q}). The uncertainty in the Q-function, which is
modeled by the covariance function in equation~(\ref{eq-Q}), gets
small as more measurements are acquired.

The parameters of the Gaussian process such as the variance $\sigma_q$ and
the kernel parameters $\sigma_f$ and $l$, can be updated at each time point using
maximum likelihood, given that the marginal likelihood of
the observed cost has the following distribution:
\beq
\c_k\mid {\bf{B}}_k\sim \N\left(\bz,\H_k(\K_k+\sigma^2_q{\bf{I}}_k\H_k^T\right))\,,
\eeq
where ${\bf{I}}_k$ is the identity matrix of size $k \times k$. For more
information, the reader is referred to~\cite{rasmussen2006gaussian}. 

\subsection{Learning the Q-function using GP-SARSA}

The Gaussian Process Temporal Difference (GPTD)
approach~\cite{engel2003bayes} is a modification of the well-known
temporal difference learning method, when the cost function over the
whole belief space is modeled by the Gaussian process and the cost is
learned based on samples of the discounted sums of returns. A {\em
  SARSA} (State-Action-Reward-State-Action) type-algorithm \cite{sutton1998reinforcement}, called
GP-SARSA, estimates the Q function using the GPTD method. 

Defining appropriate exploration/exploitation strategies for data collection has a major effect on the performance of reinforcement learning techniques. The exploration/exploitation tradeoff 
specifies the balance between the need to explore the space of all possible policies, and the necessity to focus exploitation towards policies that yield lower cost. Several policies are introduced in literature such as $\epsilon$-greedy and Boltzmann~\cite{sutton1998reinforcement}.  In this paper, 
the following policy is used for decision making~\cite{gavsic2014gaussian}:
\beq\label{eq:pi2}
\pi(\b) = \argmin_{\u\in\su} \hat{Q}(\b,\u),
\eeq
where $\hat{Q}(\b,\u)$ is a sample from $\N(\overline{Q}(\b,\u),\text{cov}((\b,\u),(\b,\u)))$, for $\u\in\su$. Notice that 
the exploration and exploitation trade-off of this policy is fully-adaptive and no parameter should be tuned. 
The GP-SARSA algorithm for control of partially-observed GRNs is
presented in Algorithm~\ref{alg:GPSARSA}. Here $\bz_{|\v|}$ denotes a
vertical vector of the same size as vector $\v$ with all elements
equal to 0.

\begin{algorithm}
\caption{\small{GP-SARSA: Control of POBDS}} 
\label{alg:GPSARSA}
\begin{algorithmic}[1]
\footnotesize
\State Initialization: $\c\lar[].$
\For{each episode}\vspace{.5ex}
\State $\b=\b_0$.\vspace{.5ex}
\If{first episode}\vspace{.5ex}
\State Select $\u\in\su$ randomly.\vspace{.5ex}
\State $\B=(\b,\u),\, {\bf{K}}\lar k((\b,\u),(\b,\u)),\,\H=[1 -\gamma]$.\vspace{.5ex}
\Else\vspace{.5ex}
\State Choose $\u\leftarrow \pi(\b)$ (Eq.~(\ref{eq:pi2})).\vspace{.5ex}
\EndIf\vspace{.5ex}

\For{each step in episode}\vspace{.8ex}
\State $\b'=\frac{T(\y)M(\u)\b}{||T(\y)M(\u)\b||_1},\,c'\lar g(\b',\u), \,\c\lar [\c\,,\,c']$.\vspace{1ex}
\If{non-terminal step}\vspace{.5ex}
\State  Choose new control $\u'\leftarrow \pi(\b')$ (Eq.~(\ref{eq:pi2})).\vspace{.5ex}
\State ${\bf{K}} \lar \begin{bmatrix}
{\bf{K}} & \k(\b',\u')\\
\k(\b',\u')^T & k((\b',\u'),(\b',\u'))
\end{bmatrix}$.\vspace{.5ex}
\State $\B\leftarrow [\B\,,\,\,(\b',\u')],\,
\H= \begin{bmatrix}
\H & \bz_{|\c|-1}\\
[\bz^T_{|\c|-1} \,\, 1] & -\gamma
\end{bmatrix}$.\vspace{.5ex}
\Else\vspace{.5ex}
\State ${\bf{H}} =\begin{bmatrix}
{\bf{H}}\\
[\bz_{|\c|-1}^T\,\,1]
\end{bmatrix}$.\vspace{.5ex}
\EndIf\vspace{.5ex}
\State Update Q-function Posterior $Q^\pi\mid\c,\B$ (Eq.~(\ref{eq-Q})).\vspace{.5ex}
\If{non-terminal step}\vspace{.5ex}
\State $\b\leftarrow \b', \u\leftarrow\u'$.\vspace{.5ex}
\EndIf\vspace{.5ex}
\EndFor

\EndFor\vspace{1ex}
\end{algorithmic}
\end{algorithm}

\section{Sparse Approximation of GP-SARSA}
\label{sec:spars}

The computational complexity of Algorithm~\ref{alg:GPSARSA} is of
order $O(k^3)$ at the time of observing the $k$th measurement. The
reason for this complexity is the need for computation of the inverse
matrix in the posterior update of GP in equation~(\ref{eq-Q}).  The
growth of this computation over time can make the GP-SARSA algorithm
computationally infeasible, especially for large POBDS, in which the
need for more data for learning Q-function seems essential.

Several techniques have been developed to limit the size of the kernel
during the learning process, such as kernel principal component
analysis (KPCA)~\cite{scholkopf1998nonlinear}, novelty criterion
(NC)~\cite{liu2009information} and the approximate linear dependence
(ALD) method~\cite{engel2003bayes}. Here, we apply the ALD method,
which constructs a dictionary of representative pairs of beliefs and
interventions online, resulting from the approximate linear dependency
condition in the feature space~\cite{engel2003bayes}.

A kernel function can be interpreted as an inner product of a set of basis functions as:
\beq
k((\b,\u),(\b,\u))\,=\, ||\Phi(\b,\u)\bullet\Phi(\b,\u)||_1\,,
\eeq
where $\bullet$ denotes the dot product of two vectors and 
\beq
\Phi(\b,\u)=[\phi_1(\b,\u),\phi_2(\b,\u),\ldots]^T\,.
\eeq

Given a set of observed beliefs and inputs
$\B=[(\b_0,\u_0),\ldots,(\b_k,\u_k)]$, any linear combination of
$\Phi(\b_0,\u_0),\ldots,\Phi(\b_k,\u_k)$ is referred to as a \textit{feature span}.  
The goal is to find the subset of points of minimum size that approximates this
kernel span. This set is called \textit{dictionary}, denoted by $D=[(\tilde{\b}_1,\tilde{\u}_1),\ldots,(\tilde{\b}_m,\tilde{\u}_m)]$ where $D\in \B$. 

The ALD condition for a new feature vector $\Phi(\b_k,\u_k)$ is:
\beq\label{eq:minPhi}
\min_{\t_k}\left|\left|\sum_{i=1}^m t_{ki}\,\Phi(\tilde{\b}_i,\tilde{\u}_i)-\Phi(\b_k,\u_k)\right|\right|^2\leq \nu
\eeq
where $(\b_k,\u_k)$ is the current point,
$\t_k=[t_{k1},\ldots,t_{km}]^T$ is the vector of coefficients, $\nu$ is
the threshold to determine the approximation accuracy and sparsity
level, and $m$ is the size of the current dictionary, $D=[(\tilde{\b}_1,\tilde{\u}_1),\ldots,(\tilde{\b}_m,\tilde{\u}_m)]$. It is shown
in~\cite{engel2005algorithms} that an equivalent minimization
to that in~(\ref{eq:minPhi}) can be written as:
\beq\label{eq:minK}
\min_{\t_k}\left( k((\b_k,\u_k),(\b_k,\u_k))-\kt_{k-1}(\b_k,\u_k)^T\,\t_k\right)\leq \nu,
\eeq
where 
\beq
\bal
\kt_{k-1}(\b_k,\u_k)=[ k((\b_k,&\u_k),(\tilde{\b}_0,\tilde{\u}_0)),\ldots,\\
& k((\b_k,\u_k),(\tilde{\b}_m,\tilde{\u}_m))]\,.
\eal
\eeq
The closed-form solution for minimization of
equation~(\ref{eq:minPhi}) is
$\t_k=\Kt_{k-1}^{-1}\,\kt_{k-1}(\b_k,\u_k)$ where $\Kt_{k-1}$ is the
Gram matrix of the points in the current dictionary.  If the threshold
in equation~(\ref{eq:minK}) exceeds $\nu$, then $(\b_k,\u_k)$ is added
to the dictionary, otherwise the dictionary stays the same.

The exact Gram matrix can be represented by:
\beq
  {\bf{K}}_k=\Phib_k^T\,\Phib_k,
\eeq
where $\Phib_k=[\Phi(\b_0,\u_0),\ldots,\Phi(\b_k,\u_k)]$.
The feature functions are approximated as $\Phi(\b_i,\u_i)\approx \sum_{j=1}^m t_{ij} \Phi(\tilde{\b}_j,\tilde{\u}_j)$, for $i=0,\ldots,k$. Defining the coefficients in a single matrix as $\T_k=[\t_0,\ldots,\t_k]^T$, we have: 
\beq\label{eq:Kt}
\bal
{\bf{K}}_k=\Phi_k^T\,\Phi_k\approx{\T}_k \tilde{\bf{K}}_k \T^T_k,\\
{\bf{k}}_k(\b_k,\u_k)\approx\T_k \tilde{\bf{k}}_k(\b,\u).
\eal
\eeq
Using equation (\ref{eq:Kt}), equation~(\ref{eq-Q}) can be approximated as:
\beq\label{eq-Q2}
Q^\pi(\b,\u)\mid {\bf{c}}_k,{\bf{B}}_k \sim \N\left(\widetilde{\Qb}(\b,\u),\widetilde{\text{cov}}\left((\b,\u),(\b,\u)\right)\right),
\eeq
where
\beq
\bal
& \widetilde{\Qb}(\b,\u) \,=\, \kt_k(\b,\u)^T \Ht_k^T(\Ht_k \Kt_k \Ht_k^T\!+\!\sigma_q^2 \Ht_k \Ht_k^T)^{-1}\c_k,\\
& \widetilde{\text{cov}}((\b,\u),(\b,\u)) \,=\, k((\b,\u),(\b,\u))\!-\!\kt_k(\b,\u)^T
\Ht_k^T\\
&\hspace{2ex}(\Ht_k\Kt_k\,\Ht_k^T+\sigma_q^2 \Ht_k \Ht_k^T)^{-1} \Ht_k \kt_k(\b,\u),
\eal
\eeq
where $\Ht_k=\H_k\T_k$. 
Using this sparsification approach allows observations to be processed
sequentially and reduces the complexity of
Algorithm~\ref{alg:GPSARSA} from $O(k^3)$ to $O(k m^2)$ where $m$ is
usually much smaller than $k$ in practice. The reader is referred
to~\cite{engel2005reinforcement} for more details. The full process of
sparsification of the GP-SARSA algorithm for learning the cost
function of POBDS is presented in Algorithm~\ref{alg:spars}. 

\begin{algorithm}[ht!]
\caption{\small{SGP-SARSA: Sparse approximation of GP-SARSA for control of POBDS}} 
\label{alg:spars}
\begin{algorithmic}[1]
\footnotesize
\State Initialize: $\alp\lar[],\,\Rt\lar[],\,\tr\lar[],\,s\lar 0,\,\frac{1}{\nu}\lar 0$.
\For{each episode}\vspace{.5ex}

\If{first episode}\vspace{.5ex}
\State Select $\u\in\su$ randomly.\vspace{.5ex}
\State $\D=\{(\b,\u)\}$, $\Kt=1/k((\b,\u),(\b,\u))$.
\Else\vspace{.5ex}
\State Choose $\u\leftarrow \pi(\b)$ (Eq.~(\ref{eq:pi2})).\vspace{.5ex}
\EndIf\vspace{.5ex}

\State $\tilde{\bf{r}}=\bz,\, s \lar 0,\,\frac{1}{\nu}\lar 0$.\vspace{.5ex}
\State $\t\lar \Kt^{-1}\kt(\b,\u),\delta\lar k((\b,\u),(\b,\u))-\kt(\b,\u)^T\t$.\vspace{.5ex}

\If{$\delta>\nu$}\vspace{.5ex}
\State $\D\lar \{(\b,\u)\}\cup \D,\,\Kt^{-1}\leftarrow \frac{1}{\delta} \begin{bmatrix}
\delta\Kt^{-1}+\t\t^T & -\t\\
-\t^T& 1
\end{bmatrix} $.\vspace{.5ex}
\State $\t\lar [\bz^T,1]^T\,,\alp\lar
 \begin{bmatrix}
\alp\\
0
\end{bmatrix},\,
\Rt\lar
 \begin{bmatrix}
\Rt & \bz\\
\bz^T & 0
\end{bmatrix},\,
\tr\leftarrow\begin{bmatrix}
\tr\\
0
\end{bmatrix}
$.\vspace{.5ex}
\EndIf\vspace{.5ex}

\For{each step in episode}\vspace{.8ex}
\State $\b'=\frac{T(\y)M(\u)\b}{||T(\y)M(\u)\b||_1},\,c'\lar  g(\b',\u)$.\vspace{1ex}
\If{non-terminal step}\vspace{.5ex}
\State \hspace{-1ex}Choose new control $\u'\leftarrow \pi(\b')$ (Eq.~(\ref{eq:pi2})).\vspace{.5ex}
\State $\t'\leftarrow \Kt^{-1}\kt(\b,\u)$.\vspace{.5ex}
\State $\delta\leftarrow k((\b',\u'),(\b',\u'))-\kt(\b',\u')^T\t'$.\vspace{.5ex}
\State $\Delta \kt\lar\kt(\b,\u)-\gamma \kt(\b',\u')$.\vspace{.5ex}
\Else\vspace{.5ex}
\State $\t'\lar \bz,\,\delta \lar 0,\,\Delta\kt\lar\kt(\b,\u)$.\vspace{.5ex}
\EndIf\vspace{.5ex}
\State $s\leftarrow \frac{\gamma\sigma_q^2}{\nu}s+c'-\Delta\kt^T\alp$. \vspace{.5ex}

\If{$\delta>\nu$}\vspace{.5ex}
\State $\D\lar \{(\b',\u')\}\cup \D$.\vspace{1ex}
\State $\Kt^{-1}\leftarrow \frac{1}{\delta} \begin{bmatrix}
\delta\Kt^{-1}+\t\t^T & -\t\\
-\t^T& 1
\end{bmatrix}$. \vspace{1ex}
\State $\t'\leftarrow [\bz^T,1]^T\,, \h\leftarrow[\t^T, -\gamma]^T$.\vspace{.5ex}
\State $\Delta k_{kk}\lar\begin{aligned}[t]
\t^T(\kt(\b,\u)-2\gamma&\kt(\b',\u'))\\
&+\gamma^2 k((\b',\u'),(\b',\u')).
\end{aligned}$\vspace{.5ex}
\State $\tr'\leftarrow \frac{\gamma\sigma_q^2}{\nu}
\begin{bmatrix}
\tilde{\bf{r}}\\
0
\end{bmatrix}+\h-
 \begin{bmatrix}
\Rt\Delta\kt\\
0 
\end{bmatrix}$.
\State $\nu\leftarrow \begin{aligned}[t]
(1+\gamma^2)\sigma_q^2+\Delta k_{kk}-\Delta \kt^T\tilde{\bf{R}}\Delta\kt&+\frac{2\gamma\sigma^2_q}{\nu}\tilde{\bf{r}}\Delta\kt
\\
&-\frac{\gamma^2\sigma_q^4}{\nu}.
\end{aligned}$
\State $\alp\lar \begin{bmatrix}
\alp\\
0
\end{bmatrix}\,,\Rt \lar \begin{bmatrix}
\Rt & \bz\\
\bz^T & 0
\end{bmatrix}$.\vspace{.5ex}
\Else\vspace{.5ex}
\State $\h\lar \t-\gamma\t',\,\tr'\lar\frac{\gamma\sigma_q^2}{\nu}\tr+\h-\Rt\Delta\kt$.\vspace{.5ex}
\If{non-terminal step}\vspace{.5ex}
\State $\nu\lar (1+\gamma^2)\sigma_q^2+\Delta\kt^T(\tr'+\frac{\gamma\sigma_q^2}{\nu}\tr)-\frac{\gamma^2\sigma_q^4}{\nu}$.\vspace{.5ex}
\Else\vspace{.5ex}
\State $\nu\lar \sigma_q^2+\Delta\kt^T(\tr'+\frac{\gamma\sigma_q^2}{\nu}\tr)-\frac{\gamma^2\sigma_q^4}{\nu}$\vspace{.5ex}
\EndIf\vspace{.5ex}

\EndIf\vspace{.5ex}
\State $\alp\lar\alp+\frac{\tr}{\nu}s,\,\Rt\lar\Rt+\frac{1}{\nu}\tr'\tr'^T,\,\tr\lar\tr',\,\t\lar\t'$.\vspace{.5ex}
\If{non-terminal step}\vspace{.5ex}
\State $\b\leftarrow \b', \u\leftarrow\u'$.\vspace{.5ex}
\EndIf\vspace{.5ex}
\EndFor
\EndFor\vspace{0ex}
\end{algorithmic}
\end{algorithm}

\section{Numerical Experiments}
\label{sec:NE}

\begin{figure}[ht!]
\begin{center}
\includegraphics[width=50mm]{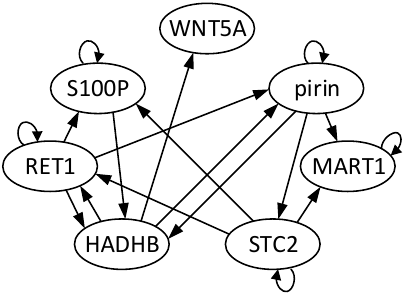}
\caption{Melanoma Gene Regulatory Network.}
 \label{fig:mal}    
\end{center} 
\end{figure}

In this section, we conduct numerical experiments using a Boolean
gene regulatory network involved in metastatic melanoma
\cite{dougherty2010stationary}. The network contains 7 genes: WNT5A,
pirin, S100P, RET1, MART1, HADHB and STC2. The regulatory relationship for this network is shown in Fig.~\ref{fig:mal} and Boolean function is presented in Table~\ref{table:WNT5A}. The $i$th output binary string specifies the
output value for $i$th input gene(s) in binary representation. For
example, the last row of Table~\ref{table:WNT5A} specifies the value
of STC2 at the current time step $k$ from different pairs of (pirin,STC2)
values at the previous time step $k-1$:\\[5pt]
(pirin=0, STC2=0)$_{k-1}$ $\rightarrow$ STC2$_k$=1\\
(pirin=0, STC2=1)$_{k-1}$ $\rightarrow$ STC2$_k$=1\\
(pirin=1, STC2=0)$_{k-1}$ $\rightarrow$ STC2$_k$=0\\
(pirin=1, STC2=1)$_{k-1}$ $\rightarrow$ STC2$_k$=1\\[5pt]

In the study conducted in~\cite{bittner2000molecular}, the expression
of WNT5A was found to be a highly
discriminating difference between cells with properties typically
associated with high metastatic competence versus those with low
metastatic competence.  Furthermore, the result of the study presented in
\cite{weeraratna2002wnt5a} suggests to reduce the activation of WNT5A
indirectly through control of other genes' activities. The reason is
that an intervention that blocked the WNT5A protein from activating
its receptor, could substantially reduce WNT5A's ability to induce a
metastatic phenotype.  For more information about the biological
rationale for this, the reader is referred
to~\cite{dougherty2010stationary}.

\begin{table}[ht!]
\caption{Boolean functions for the melanoma Boolean network.}
\begin{center}
\begin{tabular}{lll}
\toprule
Genes & Input Gene(s) & Output\\[2pt]
\midrule
WNT5A & HADHB & 10\\[2pt]
pirin & prin, RET1,HADHB & 00010111\\[2pt]
S100P & S100P,RET1,STC2 & 10101010\\[2pt]
RET1 & RET1,HADHB,STC2 & 00001111\\[2pt]
MART1 & pirin,MART1,STC2 & 10101111\\[2pt]
HADHB & pirin,S100P,RET1 & 01110111\\[2pt]
STC2 & pirin,STC2 & 1101\\[2pt]
\midrule
\end{tabular}
\label{table:WNT5A}
 \end{center}
\end{table} 

In our experiments, the intervention is applied to either RET1 or
HADHB. Recall that the intervention has uncertainty that is
modeled by a Bernoulli distribution with parameter $q$. The cost
of control is assumed to be $1$ for any taken intervention and $0$
when there is no intervention. Since the goal of control is preventing
WNT5A gene to be upregulated, the cost function can be defined as
follows:
\beq\label{eq:cost}
\bal
c(\x^j,\u)=\left\{\begin{matrix}
5+||\u||_1 & \text{if WNT5A is 1 for state }j,\\
||\u||_1 & \text{if WNT5A is 0 for state }j.\\
\end{matrix}\right.
\eal
\eeq

Table~\ref{table:para} displays the parameter values used in the experiments. The reported results are taken over 10 different runs of system during execution each with time series of length 1000. 

\begin{table}[ht!]
\scriptsize
\begin{center}
\caption{Parameter values for numerical experiments.}
\begin{tabular}{l*{1}{cc}r}
\midrule
\textbf{Parameter} & \hspace{-.8cm}\textbf{Value}\\
\midrule
\text{Number of genes $d$}  & \hspace{-.8cm} 7 \\
\midrule
\text{Number of episodes $N_{\rm ep}$}  & \hspace{-.8cm} 1, 5, 10, 15, 20 \\
\midrule
\text{Number of steps $T$}  & \hspace{-.8cm} 1000 \\
\midrule
Transition noise intensity $p$ &\hspace{-.8cm} 0.01, 0.05 \\
\midrule
Scaling variance $\sigma_f^2$ &\hspace{-.8cm} 5 \\
\midrule
Correlation parameter $l$ &\hspace{-.8cm} 0.01, 0.1, 0.2 \\
\midrule
Noise residual $\sigma_q$ &\hspace{-.8cm} 1 \\
\midrule
\text{Intervention noise intensity $q$} &\hspace{-.8cm} 0.01, 0.1, 0.2, 0.3, 0.4, 0.5 \\[2pt]
\midrule  
Initial belief $\b_0(i)$, $i=1,\ldots,128$ &\hspace{-.8cm} 1/128 \\
\midrule  
Mean in inactivated state $\mu_j^0,j=1,\ldots,7$ &\hspace{-.8cm} 40\\
\midrule  
Mean in activated state $\mu_j^1,j=1,\ldots,7$ &\hspace{-.8cm} 60\\
\midrule
\text{Standard deviations in inactivated state $\sigma_j^0$} & \hspace{-.8cm} 10, 15 \\[2pt]
\midrule
\text{Standard deviations in activated state $\sigma_j^1$} & \hspace{-.8cm} 10, 15 \\[2pt]
\midrule
\text{Discount factor $\gamma$} &\hspace{-.8cm} 0.95 \\[2pt]
\midrule
\text{Control genes} &\hspace{-.8cm} RET1, HADHB\\[2pt]
\midrule
\text{Cost function} &\hspace{-.8cm} Equation~(\ref{eq:cost})\\[2pt]
\midrule
\text{Sparcification threshold $\nu$} &\hspace{-.8cm} 0.1, 1 \\[2pt]
\midrule
\text{Value Iteration threshold $\beta$~\cite{ImanBrag:16a}} &\hspace{-.8cm} $10^{-8}$\\[2pt]
\midrule
\end{tabular}
\label{table:para}
 \end{center}
\end{table}

\subsection{Effect of GP Parameters on
the Performance of the SGP-SARSA Algorithm}
\label{Sec-exp1}

\begin{figure}[ht!]
\begin{center}
\includegraphics[width=80mm]{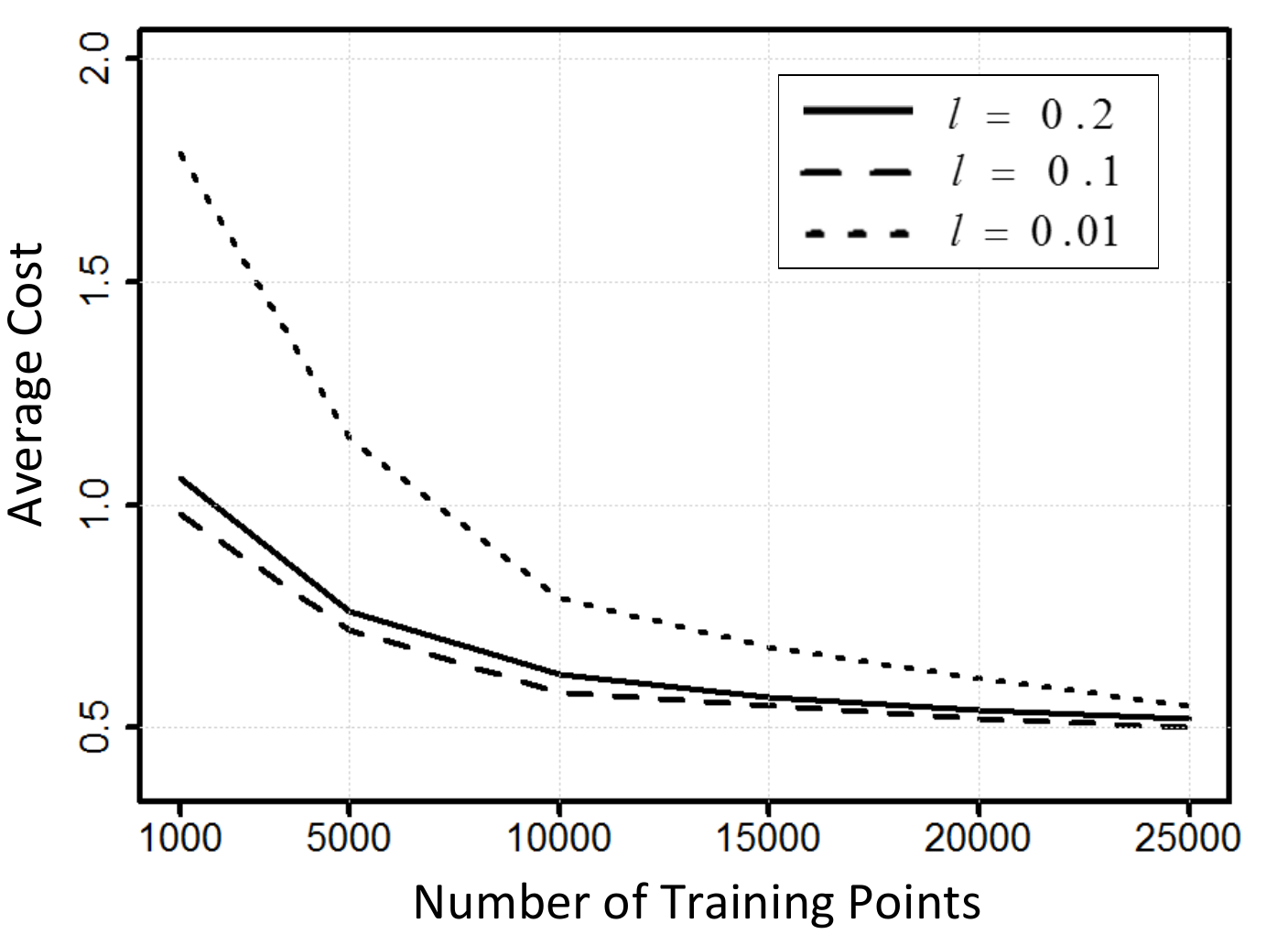}
\caption{Average cost per step achieved by SGP-SARSA as a function of the correlation coefficient.}
 \label{fig:l}    
\end{center} 
\end{figure}

In the experiments of Sections~\ref{Sec-exp1}--\ref{Sec-exp3}, RET1 is
used as the control gene and the parameters are set as follows:
$p=0.01,\, q=0.01,\, \nu=0.1,\, N_{\rm ep}=10,\, T=1000,
\,\sigma_j^0=\sigma_j^1=10,\, l=0.1,\, \nu=0.1$.
Fig.~\ref{fig:l} displays the average cost of the system under control
of SGP-SARSA for different correlation parameters $l$. The horizontal
axis shows the number of training points used in the learning process
before starting execution. It is clear that $l=0.1$ has the lowest
cost for different number of training points. In addition, $l=0.2$ has
similar cost as $l=0.1$, while $l=0.01$ behaves poorly for small
number of training points and converges to the others as the number of
training points increases. Overall, we conclude that the correlation
coefficient does not greatly affect the resulting policy, and it only
influences the speed of learning.

\subsection{Effect of Sparsification Parameter on
the Performance of the SGP-SARSA Algorithm}
\label{Sec-exp2}

\begin{figure*}
\begin{center}
\includegraphics[width=165mm]{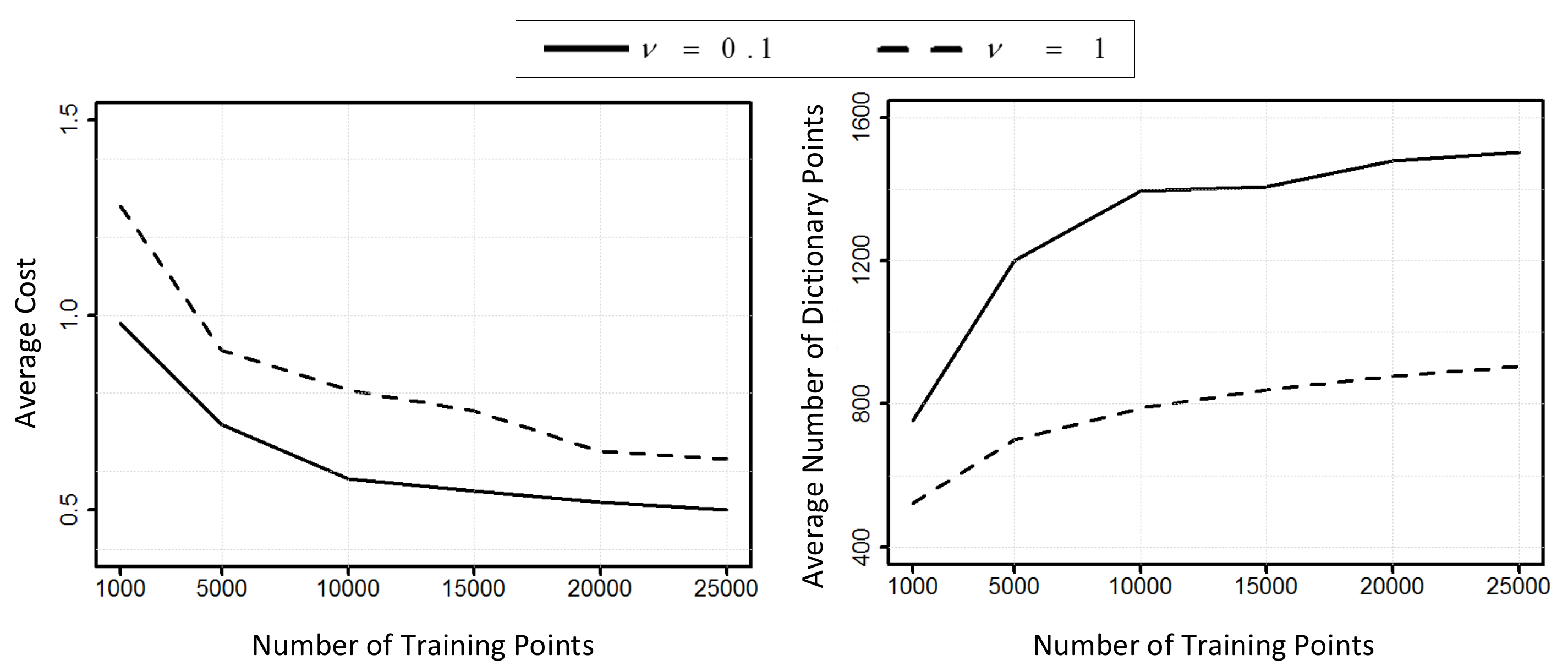}
\caption{Effect of the choice of sparsification parameter and
  dictionary size on the performance of SGP-SARSA.}
 \label{fig:nu}    
\end{center} 
\end{figure*}

Fig.~\ref{fig:nu} displays the effect of the sparsification parameter
$\nu$ on the performance of control. The right plot shows the increase
in the average number of points kept in dictionary as parameter $\nu$
gets smaller. The effect of large dictionary size can be clearly seen
in the average cost presented in left plot in Fig.~\ref{fig:nu}, in
which lower cost is achieved on average for smaller $\nu$ for
different training points.

\subsection{Effect of Transition and Intervention Noise on
the Performance of the SGP-SARSA Algorithm}
\label{Sec-exp3}

\begin{figure}
\begin{center}
\includegraphics[width=80mm]{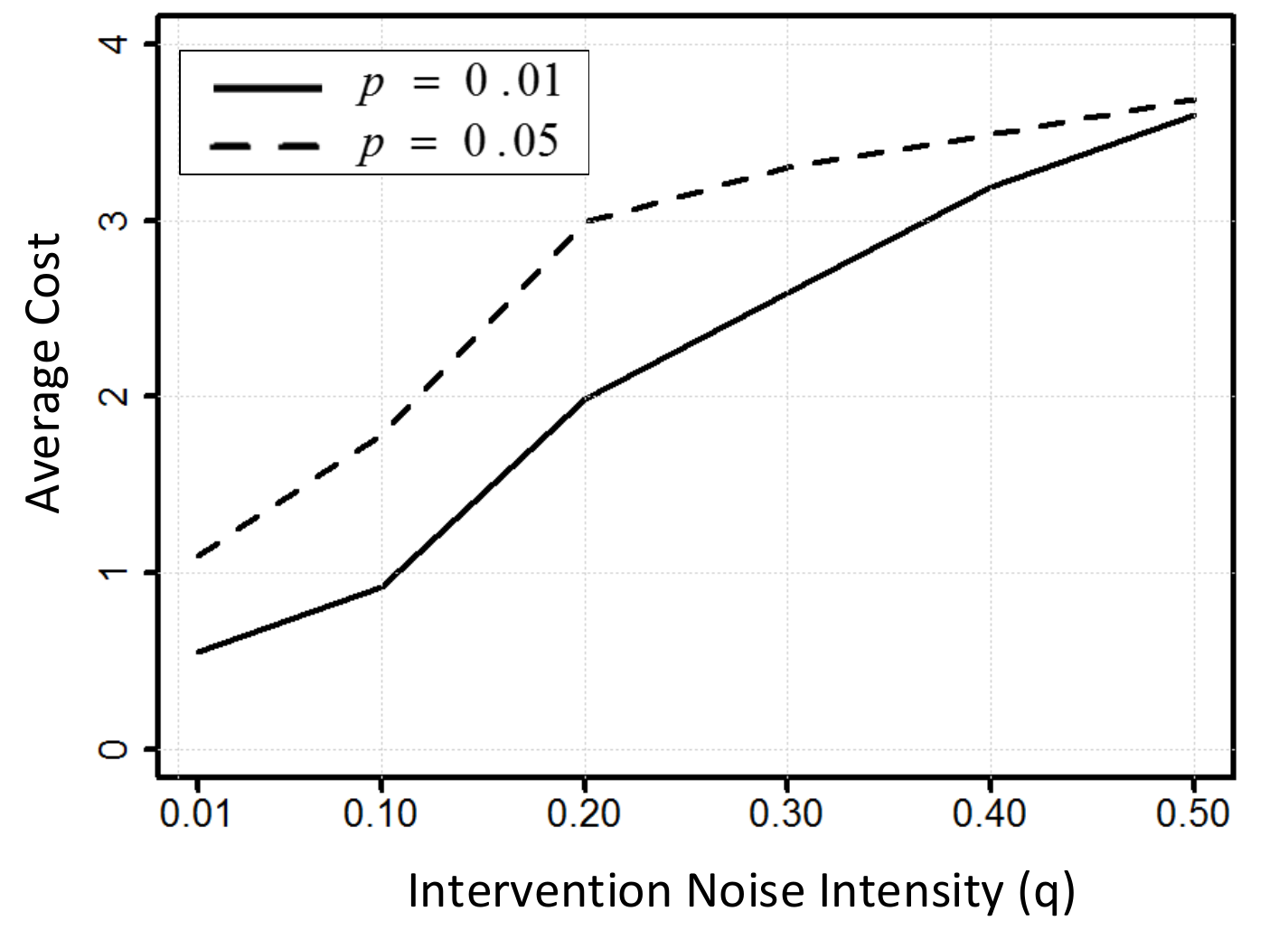}
\caption{Average cost per step achieved by SGP-SARSA as a function of
  the transition and intervention noise level.}
\label{fig:pq}    
\end{center} 
\end{figure}

Fig.~\ref{fig:pq} displays the performance of control for various
process and intervention noise levels. It is clear that the average
cost increases as the uncertainty in transition and intervention
increases, as expected. For the system without control, the average
cost is almost 2.65. By comparing this to the curves in
Fig.~\ref{fig:pq}, we reach the interesting conclusion that high
uncertainty in the intervention process can make the situation worse
than no control condition. 

\subsection{Distribution of Visited States for System under Control by
  SGP-SARSA Algorithm and without Control}
\label{Sec-exp4}

\begin{figure}[ht!]
\begin{center}
\includegraphics[width=85mm]{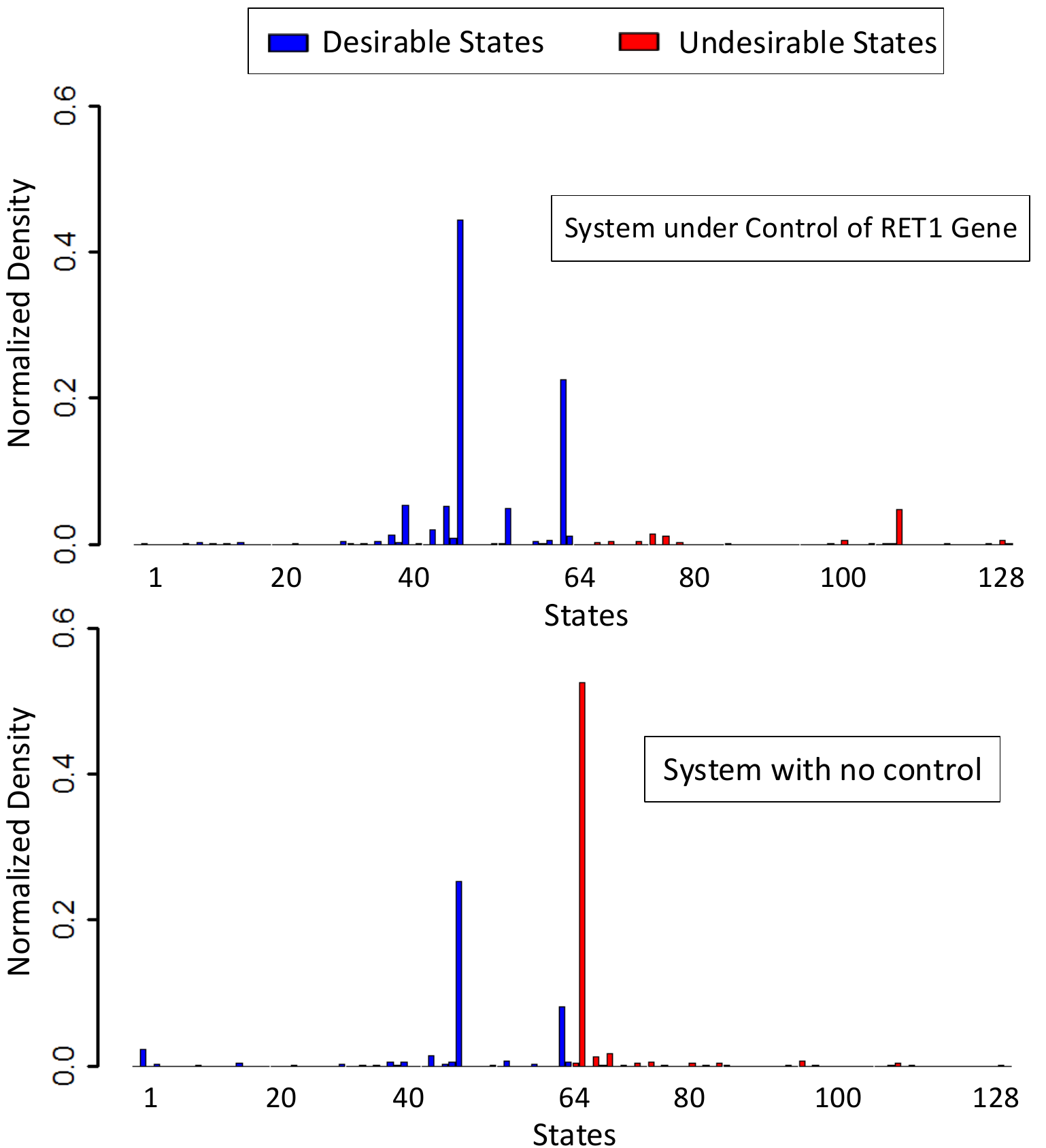}
\caption{Relative frequency of visited states under the control policy
obtained by SGP-SARSA and under no control. Desirable (inactive WNT5A)
and undesirable (active WNT5A) states are indicated by blue and red
colors, respectively.}
 \label{fig:density}    
\end{center} 
\end{figure}

Here we assess the probability mass over visited states for systems
with and without control. Fig.~\ref{fig:density} displays the long-run
relative frequencies of visited states under the control policy
obtained by SGP-SARSA and under no control. Desirable (inactive WNT5A)
and undesirable (active WNT5A) states are indicated by blue and red
colors, respectively. We can observe that the control policy obtained
by SGP-SARSA is able to shift the probability mass of visited states
from undesirable to desirable states.

\subsection{Comparison of Performance of V\_BKF and Q\_MDP Algorithms
  against the SGP-SARSA Algorithm}
\label{Sec-exp5}

Finally, we compare the performance of SGP-SARSA with two
state-feedback controllers V\_BKF~\cite{ImanBrag:16a} and
Q\_MDP~\cite{Littman95}. The intensity of uncertainty of intervention
process is set to be $q=0.1$.  The average cost per step and the
fraction of observed desirable states in the long run for the three
algorithms, for different process and observation noise levels, are
presented in Table~\ref{table:res}.

We can observe that SGP-SARSA obtains lower average cost per step than
Q\_MDP and V\_BKF, especially in the presence of high measurement
noise.  The reason is that the underlying Boolean dynamical system is
less identifiable in the presence of noisy measurements, and
therefore, the policies obtained by Q\_MDP and V\_BKF, which are not
based on the belief space but solely on the results of estimation of
the underlying Boolean dynamical system, become less valid. In
addition, we observe that RET1 is a better control input in comparison
to HADHB for reducing the activation of WNT5A in all cases.

\begin{table}[ht!]
\scriptsize
\caption{Average cost per step and average fraction of desirable
  states visited for different controllers and control genes.}
\begin{center}
\begin{tabular}{cccccccc}
\cmidrule{4-7}
&&& \multicolumn{2}{c}{RET1} &\multicolumn{2}{c}{HADHB}\\
\midrule
$p$ & $\sigma_j^0=\sigma_j^1$ &\textbf{Method} &\textbf{Cost} & \textbf{Fraction}   &\textbf{Cost} & \textbf{Fraction}  \\
\cmidrule{1-7}
\multirow{11}{*}{0.01}&\multirow{5}{*}{10}&\multirow{1}{*}{GP-SARSA}& 0.85  & 0.84 & 1.59 & 0.69\\[2pt]
\cmidrule{3-7}
&& \multirow{1}{*}{Q\_MDP} &  0.99& 0.78 & 1.82 & 0.64 \\[2pt]
\cmidrule{3-7}
&& \multirow{1}{*}{V\_BKF} & 0.98 & 0.78 & 1.86 & 0.63\\[2pt]
\cmidrule{2-7}
&\multirow{5}{*}{15}&\multirow{1}{*}{GP-SARSA}&  1.28  & 0.74 & 1.84 & 0.64\\[2pt]
\cmidrule{3-7}
&& \multirow{1}{*}{Q\_MDP} &  1.64 & 0.67 & 2.19 & 0.56 \\[2pt]
\cmidrule{3-7}
&& \multirow{1}{*}{V\_BKF} & 1.62  & 0.67  & 2.17 & 0.57\\[2pt]
\cmidrule{1-7}
\multirow{11}{*}{0.05}&\multirow{5}{*}{10}&\multirow{1}{*}{GP-SARSA}& 1.79  & 0.64 & 2.18 & 0.58\\[2pt]
\cmidrule{3-7}
&& \multirow{1}{*}{Q\_MDP} & 2.09& 0.58 & 2.32 & 0.54 \\[2pt]
\cmidrule{3-7}
&& \multirow{1}{*}{V\_BKF} & 2.07  & 0.58 & 2.34& 0.54\\[2pt]
\cmidrule{2-7}
&\multirow{5}{*}{15}&\multirow{1}{*}{GP-SARSA}&  2.08  & 0.60 & 2.53 & 0.50\\[2pt]
\cmidrule{3-7}
&& \multirow{1}{*}{Q\_MDP} &  2.35 & 0.54 & 2.73 & 0.46 \\[2pt]
\cmidrule{3-7}
&& \multirow{1}{*}{V\_BKF} & 2.37  & 0.53  & 2.75 & 0.47\\[2pt]
\midrule
\end{tabular}
\label{table:res}
 \end{center}
\end{table}

\section{Conclusion}
\label{sec:con}

In this paper, The POBDS model was used in conjunction with Gaussian
process and reinforcement learning to achieve near-optimal
infinite-horizon control of gene regulatory networks with uncertainty
in both the inputs (intervention) and outputs (measurements).  The
cost function in the belief and intervention spaces was modeled by
Gaussian process and learning was achieved using a sparsified version
of the GP-SARSA algorithm. The methodology was investigated thoroughly
by a series of numerical experiments using synthetic gene-expression
data generated by a gene regulatory network involved in melanoma
metastasis. An interesting fact observed in the expriments is that if
the uncertainty in the control input is large, the behavior of the
controlled system is worse than that of a free-evolving system
evolving without control. Future work will consider adaptive version
of the controllers described here.

\section*{Acknowledgment}

The authors acknowledge the support of the National Science Foundation,
through NSF award CCF-1320884.

\bibliographystyle{ieeetr}
\bibliography{ref}

\end{document}